\pdfoutput=1

\documentclass[twocolumn]{aastex701}
\usepackage{newtxtext,newtxmath}
\usepackage[T1]{fontenc}

\usepackage{graphicx}	%
\usepackage{amsmath}	%

\usepackage[utf8]{inputenc}
\usepackage{indentfirst}
\usepackage{mathrsfs}
\usepackage{amsmath,siunitx}
\usepackage{bm}
\usepackage[version=4]{mhchem}

\newcommand{\adrift}{a_\mathrm{drift}}	%
\newcommand{\afrag}{a_\mathrm{frag}}	%
\newcommand{\amax}{a_\mathrm{max}}      %
\newcommand{\co}{\ce{CO}}
\newcommand{\coo}{\ce{CO2}}

\newcommand{\gw}{\Delta_{\mathrm{gap}}}
\newcommand{\gd}{\Sigma_{\rm{gap}}/\Sigma_{0}}

\newcommand{\hho}{\ce{H2O}}

\newcommand{\Msun}{M_\odot}

\newcommand{\Mj}{M_\mathrm{J}}

\newcommand{\nphotsca}{nphot_{\rm{scat}}}

\newcommand{\rp}{r_{\mathrm{p}}} %

\newcommand{\sigmad}{\Sigma_{\rm{dust}}}
\newcommand{\sigmag}{\Sigma_{\rm{gas}}}

\newcommand{\sigmagap}{\Sigma_{\rm{gap}}}

\newcommand{\Tmid}{T_{\rm{mid}}}

\newcommand{\vf}{V_\mathrm{frag}}	%
\newcommand{\w}{w}

\newcommand{\radmc}{\text{RADMC-3D}}

\defcitealias{chen2024}{C24}

\begin{document}

\title{Planet-gap-induced ice lines form observable dust rings in protoplanetary disks}

\correspondingauthor{Kan Chen}
\author{Kan Chen}
\affiliation{Kavli Institute for Astronomy and Astrophysics, Peking University, Beijing 100871, People's Republic of China}
\email[show]{kanchen@pku.edu.cn}

\author{Tommy Chi Ho Lau}
\affiliation{Department of Astronomy \& Astrophysics, University of Chicago, Chicago, IL 60637, USA}
\email{tommylauch@uchicago.edu}

\author{Paola Pinilla}
\affiliation{Mullard Space Science Laboratory, University College London, Holmbury St Mary, Dorking, Surrey RH5 6NT, UK}
\email{p.pinilla@ucl.ac.uk}

\begin{abstract}
Whether volatile ice lines can produce dust rings in protoplanetary disks
remains debated.
Previous modelling and observational studies place ice lines
using smooth, monotonic disk temperature profiles.
However, in substructured disks, the midplane temperature need not be smooth: a
deep, planetary gap can locally heat, shift and multiply the
ice-line locations.
We adopt the non-smooth temperature profile of a hot planetary
gap and use dust-evolution simulations to study how \coo\ ice lines, which
imprint radial jumps in the fragmentation velocity $\vf$, affect dust evolution
over 10~Myr and the observational signatures.
For comparison, we also run models with uniform low ($\vf=1$~m/s) and
high ($\vf=10$~m/s) fragmentation velocities.
In the varying-$\vf$ model, dust piles up with enhanced grain sizes at the two
\coo\ ice lines along the gap edges, appearing as clear double rings
in dust observations at 13~$\mu$m that persist over 10~Myr; at 1.3~mm, it
shows only a modest ring-to-gap contrast of a factor of a few, consistent with
the shallow, low-contrast rings in ALMA surveys.
The uniform $\vf=10$~m/s model instead forms a single high-contrast ring at the
pressure bump and suppresses the inward pebble flux to the inner
disk by up to two orders of magnitude, whereas the uniform $\vf=1$~m/s model
produces only weak substructure and strong outer-disk depletion.
We propose a multi-wavelength test to distinguish these scenarios: ice-line
rings stay at fixed radii from scattered light to the millimetre observation, whereas a
pressure-bump trap produces a single dominant ring whose peak shifts inward at
shorter wavelengths.

\end{abstract}


\section{Introduction}
ALMA has revealed widespread dust gaps and rings in protoplanetary disks \citep[e.g.][]{andrews2018, long2018, cieza2021}, and their origin remains debated. One proposed origin is ice lines, where a volatile freezes out and changes the local dust properties. Ice lines alter dust growth in two coupled ways. First, sublimation and recondensation change the grain mass and the local solid budget \citep{drazkowska2017,schoonenberg2017,hyodo2019}. Second, icy grains are stickier than dry ones, so the fragmentation velocity $\vf$ changes across an ice line and reshapes the grain-size and surface-density distributions \citep{banzatti2015,pinilla2017,okuzumi2019}. These effects may act together, and both can pile up dust near ice lines. Importantly, in all these models the ice line locations are set by a smooth, power-law disk temperature profile.

From observations, although \citet{zhang2015} interestingly found that the locations of dust gaps and rings in the HL~Tau disk seem to correlate with the ice line locations of some volatiles, recent ALMA surveys suggest no simple correlation between dust substructure locations and stellar luminosity \citep{huang2018,long2018}. More specifically, these surveys found that the observed dust ring locations do not align with the predicted ice line locations derived by assuming a simple smooth temperature profile based on stellar luminosity.

However, the temperature profile in disks may not be smooth, especially in disks with substructures. \citet{izquierdo2023} find temperature deviations at gaps and rings in the disks of MWC~480 and HD~163296, mostly warm gas gaps and cold dust rings. Simulations show that planet wakes can also shift ice lines directly through shock heating \citep{ziampras2020}. Recent modeling work by \citet{chen2024,chen2025} also shows that a planet-induced gap can cause an increase in the midplane temperature within the gap and a decrease in temperature at the dust rings. In their models, they find that, due to the reduced optical depth caused by density depletion in the gap, stellar photons can more easily penetrate into deeper regions and heat the disk midplane. Such non-smooth temperature profiles can further shift the locations of ice lines, meaning that they cannot be simply predicted from smooth temperature profiles based solely on stellar luminosity. Furthermore, structured temperature profiles can feed back into the evolution of gas and dust density substructures.

In this paper, we aim to study how the non-smooth temperature profile induced by a planet gap can affect volatile ice line locations and further affect dust evolution and observational signatures.
We isolate the fragmentation-velocity channel, imposing $\vf$ jumps at the ice lines without including the mass exchange from sublimation and recondensation. We study dust evolution, in particular the dust size and density distribution at 1, 5, and 10~Myr.

This paper is structured as follows. In Section~2, we describe the setup and results of our dust evolution simulations. In Section~3, we post-process the dust evolution results with radiative transfer to produce synthetic observations, with emphasis on dust rings. In Section~4, we discuss the observational implications of our synthetic signatures, pebble flux into inner disks, and the limitations of the modeling approach. Finally, we summarize our main conclusions in Section~5.

\section{Dust evolution model}
\label{sec:dust}

\subsection{Setup}

We model dust evolution including dust growth, fragmentation, radial drift, and turbulent diffusion with the 1D code \texttt{DustPy} \citep{stammler2022}. The gas surface density, midplane temperature, and gas radial and azimuthal velocity profiles ($v_r$, $v_\phi$) are held fixed in time while dust evolves; we do not include the back reaction from dust to gas. Our aim is to isolate how changes in the fragmentation velocity $\vf$ across ice lines affect the dust size and surface density distributions.

We take the gas profiles from the iterative hydrodynamical and radiative transfer simulations of \citet{chen2024}. In that work, 2D $(r,\phi)$ planet--disk models are iterated between hydrodynamics and radiative transfer to capture how gap opening modifies the thermal structure and how those changes feed back on the planet--disk interaction. They iterate the gap-opening process to 2000 planetary orbits to obtain the quasi-steady state gas profiles. We adopted their reference disk, which extends from 1 to 100~au with a mass of $0.028\,\Msun$ around a $1\,\Msun$ star. We azimuthally averaged their results of the 2D $\sigmag$, $\Tmid$ and $v_\phi$ to obtain the 1D inputs used here. The $3\,\Mj$ planet at 30~au opens a nearly circular gap, so it is reasonable to do azimuthal averaging.

\begin{figure}
    \includegraphics[width=\columnwidth]{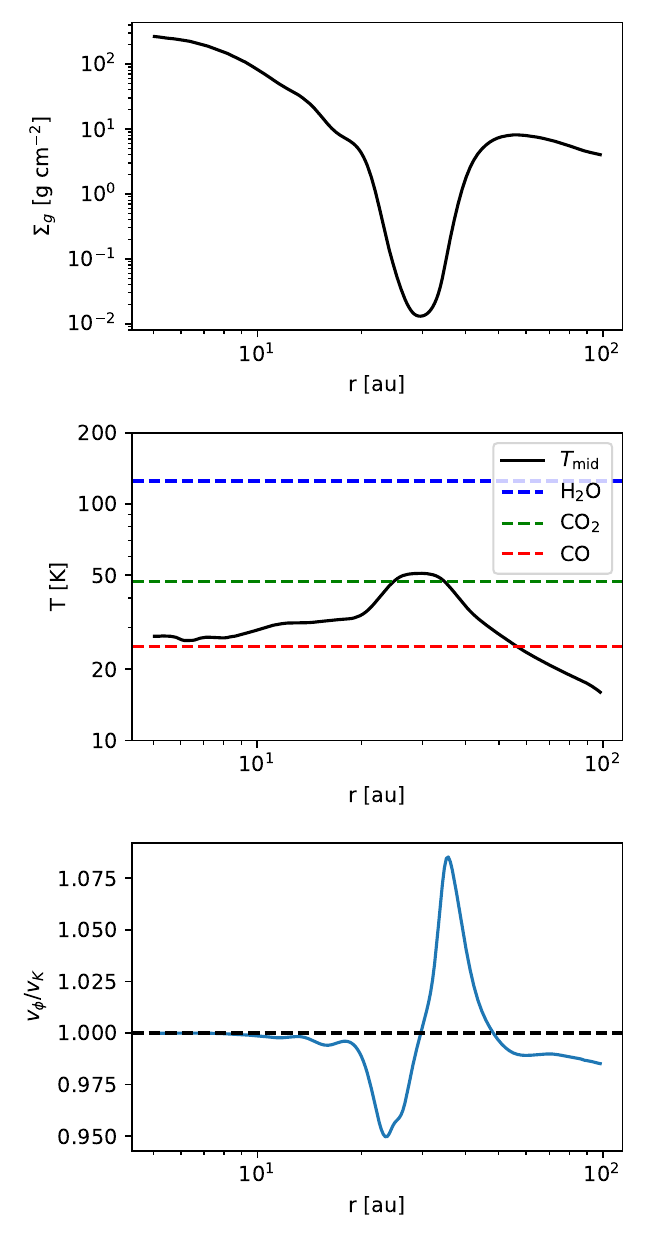}
    \caption{
    Gas profiles used as fixed inputs to the dust evolution simulations for an $\alpha=10^{-3}$ disk with a $3\,\Mj$ planet at 30~au. From top to bottom: gas surface density, midplane temperature, and gas azimuthal velocity.
    Horizontal dashed lines in the middle panel indicate the \hho, \coo, and \co\ sublimation temperatures (blue, green, and red, respectively).
    }
    \label{fig:setup_visc1e-3}
\end{figure}

We adopt the model with a $3\,\Mj$ planet at $30$~au and turbulent viscosity $\alpha=10^{-3}$, where the kinematic viscosity is $\nu=\alpha\,c_\mathrm{s}H$, with $c_\mathrm{s}$ the sound speed and $H$ the gas pressure scale height \citep{shakura1973}. This case produces a deep, wide gap (gap depth $\gd \sim 10^{-3}$, where $\sigmagap$ is the gas surface density at the gap centre and $\Sigma_0$ the unperturbed background value; full gap width $\gw \sim 0.9\,\rp$, where $\rp$ is the planet's orbital radius) with a midplane temperature increase of $\sim$20~K at the gap centre relative to the gap edges, sufficient to sublimate CO$_2$ ice locally. Note that in the \citet{chen2024} model adopted here, the elevated midplane temperature in the gap arises mainly from radiative heating: reduced dust optical depth lets stellar radiation penetrate deeper, leading to higher temperature in the midplane of the gap region. Interior to the gap, the midplane temperature is lowest near 8~au, which lies in the shadow of the puffed-up inner rim of the \citet{chen2024} disk, and rises slowly between 10 and 20~au as the disk emerges from the shadow. This rise is not caused by the planet. They do not consider the contribution from viscous heating or spiral shock heating. The azimuthal velocity panel shows a super-Keplerian to sub-Keplerian transition at the outer gap edge near 50~au, marking a pressure bump (Fig.~\ref{fig:setup_visc1e-3}). For the gas radial velocity, we use the viscous accretion velocity of the fixed $\sigmag$, the \texttt{DustPy} default \citep{stammler2022}. Appendix~\ref{sec:app-gapT} compares this model with the no-planet and $1\,\Mj$ models of \citet{chen2024}. Only the $3\,\Mj$ gap exceeds the \coo\ sublimation temperature.

\begin{figure}
    \includegraphics[width=\columnwidth]{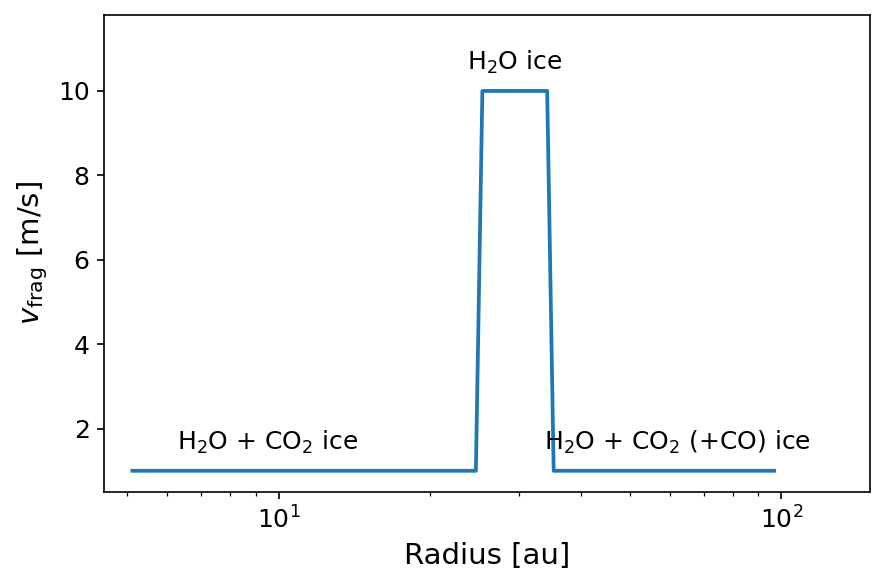}
    \caption{
    Fragmentation velocity profile used in the varying-$\vf$ model, derived from the midplane temperature profile in Fig.~\ref{fig:setup_visc1e-3}. In regions where grains are coated only with water ice, $\vf=10$~m/s; elsewhere, $\vf=1$~m/s. CO is shown in parentheses because it freezes out at $\sim$56~au but does not change $\vf$ there, since \coo\ is already frozen.
    }
    \label{fig:vfrag_r}
\end{figure}

The fragmentation velocity $V_\mathrm{frag}$ is the threshold collision velocity above which dust grains fragment rather than stick. We allow $\vf$ to change across ice lines, following \citet{pinilla2017}. We adopt sublimation temperatures of 125~K, 47~K, and 25~K for $\hho$, $\coo$, and $\co$, respectively \citep{oberg2011}. Grains coated only with water ice have $\vf=10$~m/s; grains coated with the mixture of water and other ices have $\vf=1$~m/s \citep[e.g.,][]{blum2008, wada2009, gundlach2014, musiolik2019}. Applied to the midplane temperature profile in Fig.~\ref{fig:setup_visc1e-3}, this yields $\vf$ jumps at the \coo\ ice lines near 25~au (sublimation) and 35~au (condensation; Fig.~\ref{fig:vfrag_r}).

On this same gas background, we perform three dust evolution runs: the varying-$\vf$ model, with $\vf$ jumping across ice lines, and two control runs with $\vf=1$~m/s and $\vf=10$~m/s fixed over the entire disk, mimicking cases that ignore ice-line-induced changes in $\vf$, for example if water ice loses its high stickiness at low temperatures \citep{musiolik2019}. We refer to these as the varying-$\vf$, $\vf=1$~m/s, and $\vf=10$~m/s models. The gas pressure profile is identical across all three models.

We evolve dust from 5 to 100~au with $n_r=100$ logarithmically spaced radial grid points. Tests with doubled radial resolution yield similar results. We exclude the inner 5~au because a short pure-water-ice region there lies outside our focus on ice-line effects near the planet at $\sim$30~au; tests including 1--5~au give essentially the same results for $r>5$~au. The initial dust-to-gas ratio is 0.01. Grains initially follow the MRN size-distribution from 0.1 to 1~$\mu$m with power-law index $-3.5$ \citep{mathis1977}. During evolution we track sizes from 0.1~$\mu$m to 10~cm with seven bins per decade in $\log a$, where $a$ is the dust grain size. The internal grain density is $\rho_\mathrm{s}=1.67\,\mathrm{g\,cm^{-3}}$. We remove the dust mass to transfer to planetesimal formation when dust-to-gas ratio reaches unity \citep{schoonenberg2018}. This threshold assumes that the streaming instability \citep{youdin2005,johansen2007} operates once the midplane dust-to-gas ratio reaches unity, whereas strong clumping also depends on the Stokes number \citep{lim2025}. Because planetesimal formation removes dust from the ice-line peaks, a stricter criterion would strengthen rather than erase the ring signatures discussed below.

\subsection{Results}
\label{sec:dust-results}

\begin{figure*}
	\includegraphics[width=\linewidth]{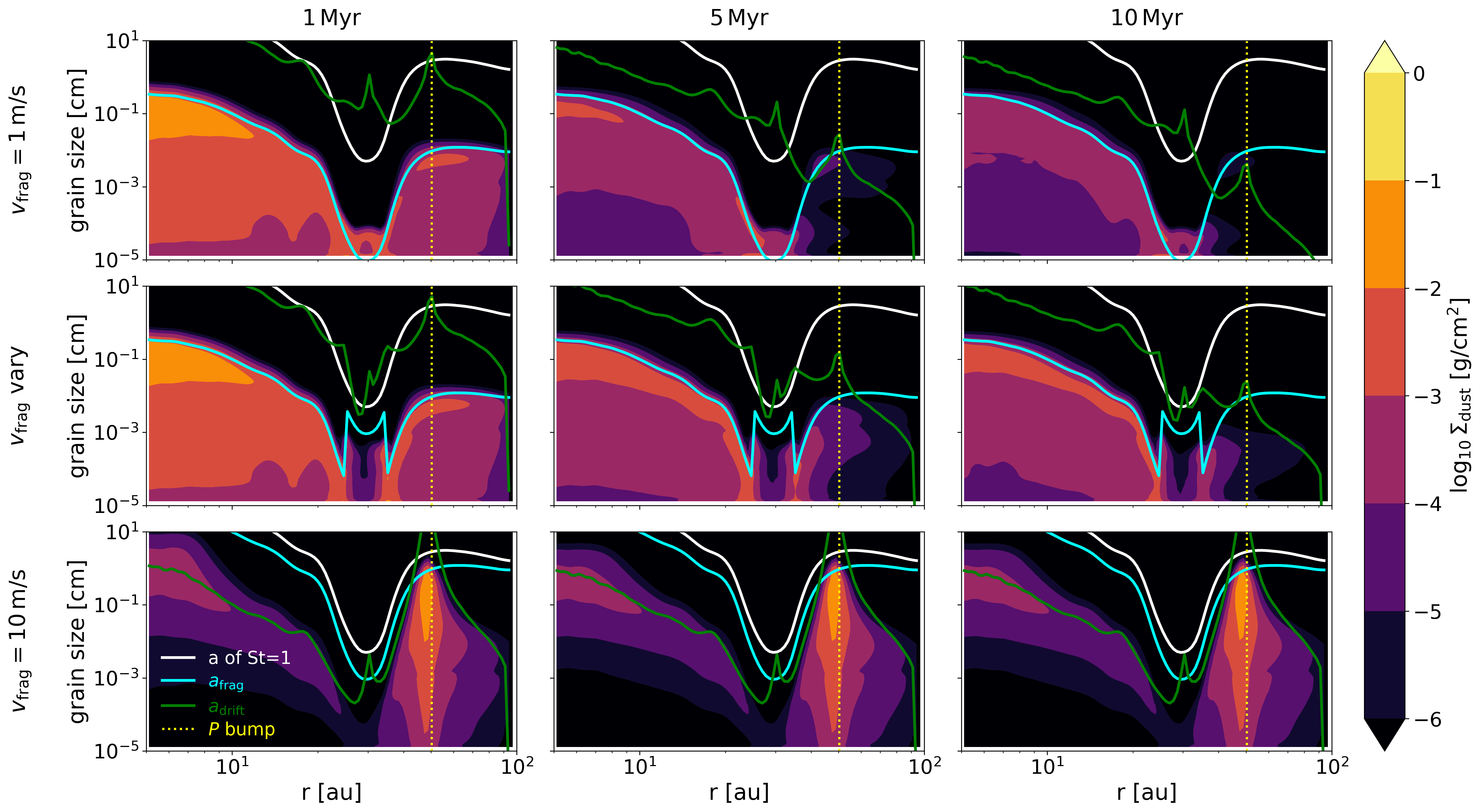}
    \caption{Vertically integrated dust density distribution at 1~Myr (left), 5~Myr (middle), and 10~Myr (right) of evolution. The top, middle, and bottom panels show the $\vf=1$~m/s, varying-$\vf$, and $\vf=10$~m/s models, respectively. The varying-$\vf$ model has a jump in $\vf$ across the \coo\ ice lines along the gap region (Fig.~\ref{fig:vfrag_r}), whereas the $\vf=1$ and $\vf=10$~m/s models use a radially constant $\vf$ over the whole disk. Cyan, green, and white lines show the fragmentation limit $\afrag$, drift limit $\adrift$, and the grain size corresponding to St\,=\,1, respectively. The yellow dotted line marks the radial location of the local gas pressure maximum (pressure bump) at the outer gap edge.
    }
    \label{fig:sdust_visc1e-3}
\end{figure*}

In Fig.~\ref{fig:sdust_visc1e-3}, we compare the $\vf=1$~m/s (top), varying-$\vf$ (middle), and $\vf=10$~m/s (bottom) models on the dust size and density distributions over 1, 5, and 10~Myr of evolution (left to right).
We also overplot the fragmentation limit $\afrag$ (cyan lines), drift limit $\adrift$ (green lines), and the grain size corresponding to St\,=\,1 (white lines) to help analyse the dust evolution.
The fragmentation limit $\afrag$ \citep{birnstiel2012} is calculated from
\begin{equation}
    \afrag = \frac{2}{3\pi} \frac{\Sigma_\mathrm{g}}{\rho_\mathrm{s}\alpha} \frac{\vf^2}{c_\mathrm{s}^2},   \label{eq:afrag}
\end{equation}
where $\Sigma_\mathrm{g}$ is the gas surface density, $\alpha$ is the turbulence parameter, and $c_\mathrm{s}$ is the sound speed.

The drift limit $\adrift$ \citep{birnstiel2012} is calculated by:
\begin{equation}
    \adrift = \frac{2\Sigma_\mathrm{d}}{\pi\rho_\mathrm{s}} \frac{v_\mathrm{k}^2}{c_\mathrm{s}^2} \left| \frac{d \ln P}{d \ln r} \right|^{-1},
\end{equation}
where $\Sigma_\mathrm{d}$ is the dust surface density, $v_\mathrm{k}$ is the Keplerian velocity, and $P$ is the gas pressure.

The times in Fig.~\ref{fig:sdust_visc1e-3} are the duration of the dust evolution on a fixed gas background, not disk ages, since our calculation starts from an already-formed gap. In reality the gap opens over a finite time and the dust evolves alongside the gas, but we hold the gas fixed to isolate the effect of $\vf$ changes across ice lines.

\subsubsection{Varying-$\vf$ model}

In the varying-$\vf$ model (middle row of Fig.~\ref{fig:sdust_visc1e-3}), at 1~Myr, $\amax$ shows distinct jumps at the $\coo$ ice line locations. Across the whole disk, $\amax$ is limited by fragmentation. In the region between the $\coo$ ice lines (25--35~au), where grains are coated with only water ice, $\vf$ increases to $\vf=10$~m/s, allowing grains to grow larger. This results in $\amax$ of about 10~$\mu$m at both of the ice lines. In regions immediately adjacent to the ice lines, $\amax$ decreases to 1~$\mu$m because $\vf$ drops to $\vf=1$~m/s. However, we can also note that the $\amax$ is still about a few times smaller than $\afrag$ within the gap region. This is mainly because the gap region is a low-density region, and the grains have incomplete collisional growth. Also, at the two ice lines, the dust-to-gas ratio is reaching unity, which means some grains have been converted to planetesimals and removed from the dust size bins. This conversion continues until $\sim$5~Myr, when it saturates, removing $\sim$$40\,M_\oplus$ of solids in total, about half of the initial dust mass.
In short, the enhancement in grain size at the gap region highlights the influence of ice-line-induced changes in $\vf$ on dust evolution.

By 1~Myr, the surface density in the varying-$\vf$ model in the inner and outer disk is similar to that in the $\vf=1$~m/s model because both models in these regions have $\vf=1$~m/s. Within the gap region, at the two jumps in $\afrag$ caused by the ice lines, we also observe corresponding jumps in surface density compared to the regions immediately outside these jumps. For the gap center region, due to the low gas surface density, small grains also have low density. Therefore, we form double density peaks at the two ice lines.

By 5~Myr, the distinct jumps in $\amax \sim 10\,\mu$m at the ice-line locations persist, but the outer disk becomes more depleted due to radial drift, with $\sigmad$ at 60~au falling by a factor of $\sim$70 from 1~Myr. The corresponding jumps in $\sigmad$ at the two $\afrag$ transitions associated with the ice lines remain visible.

By 10~Myr, the varying-$\vf$ model continues to lose dust in the outer disk through radial drift, leading to an overall lower $\sigmad$ at large radii (a further factor of $\sim$5 at 60~au). The $\amax \sim 10\,\mu$m transitions at the ice-line locations remain visible as sharp jumps in $\afrag$, and the associated local changes in $\sigmad$ persist near these radii.

\subsubsection{$\vf=1$~m/s model}

In Fig.~\ref{fig:sdust_visc1e-3}, at 1~Myr, in the $\vf=1$~m/s model (top row), the maximum grain size $\amax$ is primarily limited by the fragmentation limit ($\afrag$) throughout the disk. As $\afrag$ is proportional to $\sigmag$ in Eq.~\ref{eq:afrag}, the shape of $\afrag$ also generally follows the shape of the gas surface density. The low $\vf$ restricts grain growth; at the same time, the hot gap region yields a very low $\afrag$ there, resulting in small $\amax$ of about 1~$\mu$m at the gap. We note that $\afrag$ drops near our minimum grain size in the depleted gap. The dust distribution extends slightly above the cyan line because Eq.~\ref{eq:afrag} is a diagnostic limit, not an exact upper bound on the simulated distribution.

In the same 1~Myr, $\vf=1$~m/s panel (top-left of Fig.~\ref{fig:sdust_visc1e-3}), the grain surface density distribution shows that larger grains (e.g., 0.1~mm to 1~mm) are more concentrated in the inner disk (inside 20~au), leading to a higher surface density in this region. In the outer disk (outside 40~au), the surface density of grains is relatively low due to the combined effects of radial drift and slower grain growth.
Within the gap region, the surface density is significantly reduced for almost all grain sizes due to the planet-induced gas gap. However, small grains ($<1\,\mu$m) are less affected by radial drift and maintain a relatively high surface density by turbulence diffusion.

By 5~Myr, within 50~au, $\amax$ remains limited by $\afrag$ and is similar to the 1~Myr result. However, in drift-limited regions outside 50~au, grains grow and drift inward until the outer grain population is largely depleted in $\sigmad$ (by a factor of $\sim$300 at 60~au). The dust surface density also decreases by a factor of $\sim$20 in the inner disk and $\sim$5 in the gap compared with 1~Myr.

By 10~Myr, $\amax$ in the outer disk is broadly similar to that at 5~Myr, while the already depleted outer-disk $\sigmad$ drops by a further factor of $\sim$10. The inner-disk $\sigmad$ also continues to decrease due to radial drift, by a factor of $\sim$2 at 10~au.

\subsubsection{$\vf=10$~m/s model}

In the $\vf=10$~m/s model at 1~Myr (bottom row of Fig.~\ref{fig:sdust_visc1e-3}), $\amax$ is primarily limited by $\adrift$ in most regions of the disk. The high $\vf$ allows grains to grow larger, and radial drift becomes the dominant mechanism restricting further growth. Near the pressure bump at the outer gap edge, $\amax$ is closer to $\afrag$, as the pressure bump traps dust grains and enhances their growth. This results in larger grains (up to about 1~cm) and a peak in $\sigmad$ (reaching about 0.1~g~cm$^{-2}$ for 0.1~mm to 1~cm grains) at the trap, while the outer disk remains drift limited. The peaks in $\amax$ and $\sigmad$ are both located slightly interior to (at smaller $r$ than) the yellow dotted line marking the local gas pressure maximum: because $\afrag$ still rises with $r$ approaching the bump, the fragmentation limit caps grain growth, and hence dust accumulation, just inside the exact pressure maximum rather than at its peak.

By 5~Myr, the $\amax$ and $\sigmad$ are relatively similar to the 1~Myr result ($\sigmad$ at the dust trap changes by only a few per cent). Strong dust trapping at the pressure bump stops radial drift and retains large grains in the outer disk. Meanwhile, turbulent diffusion allows some small grains to leak through the gap into the inner disk, helping to maintain $\amax$ and $\sigmad$ there.

By 10~Myr, the $\amax$ and surface density distribution in the outer disk remain broadly similar to those at 5~Myr, indicating that dust evolution has largely reached a quasi-steady state under the fixed gas profiles.

\section{Synthetic observations}
\label{sec:rt}

\subsection{Setup}
\label{sec:rt-setup}

We post-process the dust evolution results from all three models (varying-$\vf$, $\vf=1$~m/s, and $\vf=10$~m/s) at 1, 5, and 10~Myr with the radiative transfer code $\radmc$ \citep{dullemond2012} to produce synthetic observations. We use the same radial grid as in the dust evolution simulations and adopt the colatitude grid from \citet{chen2024}  hydrodynamical and RT model ($n_\theta=53$ cells from the midplane to the upper disk surface, $\theta=\pi/2-0.5$ to $\pi/2$, with mirror symmetry about the equatorial plane). The disk is assumed to be axisymmetric in the azimuthal direction.

For the grain properties, we use the DSHARP opacity \citep{birnstiel2018} in $\radmc$. The DustPy size distribution is resampled into 16 species logarithmically spaced in size between 0.1~$\mu$m and 10~cm for the RT calculation.

\begin{figure*}
	\includegraphics[width=0.9\linewidth]{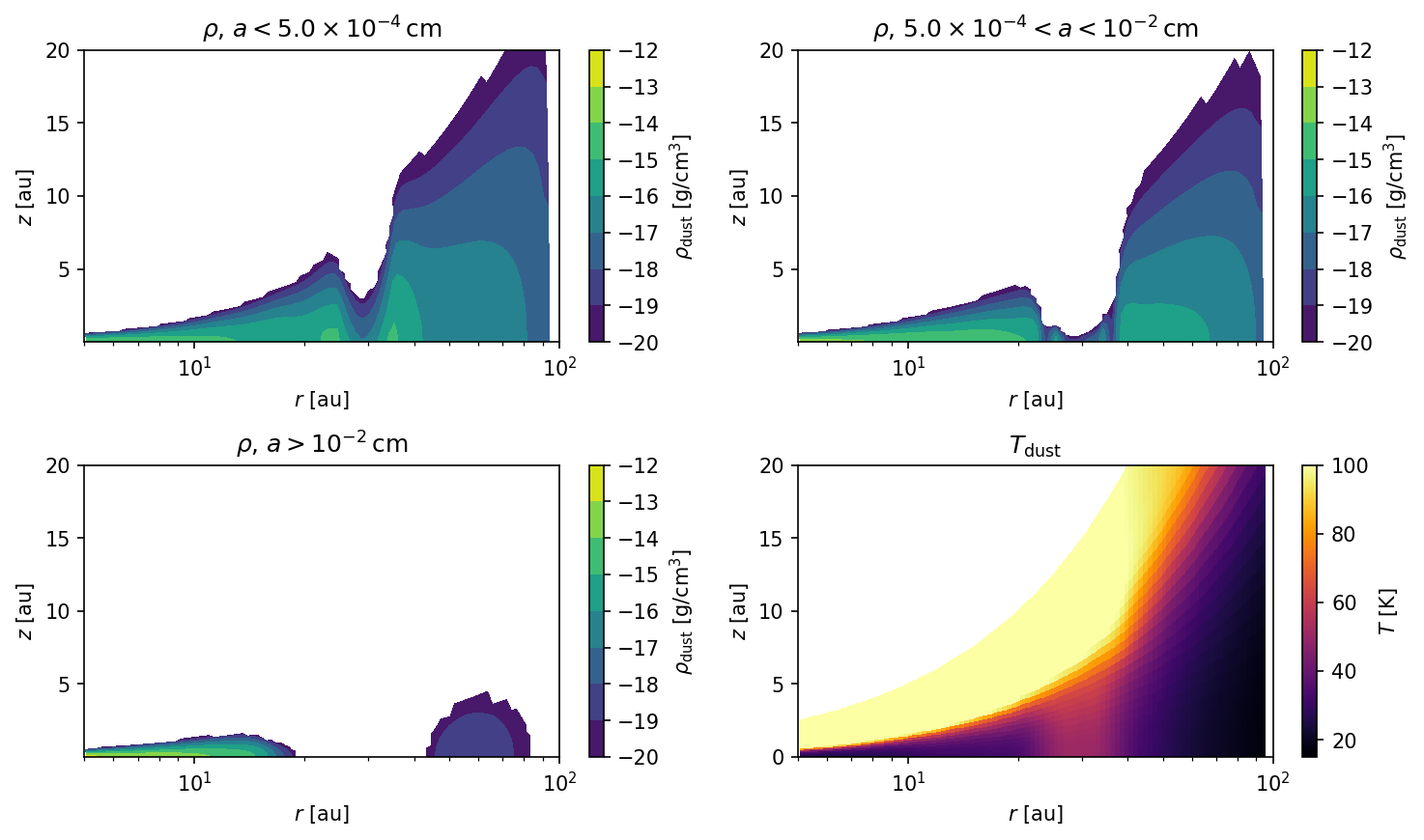}
    \caption{
    Dust volume density distribution of small ($<5\,\mu$m; upper left), medium (5--100~$\mu$m; upper right), and large ($>100\,\mu$m; lower left) grains in the varying-$\vf$ model at 1~Myr of evolution. The lower right panel shows the 2D $(r,\theta)$ temperature structure used in $\radmc$ for synthetic observations. For display purposes, we do not show grains with densities lower than $10^{-20}$~g~cm$^{-3}$ in the dust density panels.
    }
    \label{fig:radmc_setup_1myr}
\end{figure*}

We extend the 1D radial dust surface density profile to 3D in spherical coordinates. The dust vertical distribution is calculated by assuming vertical hydrostatic equilibrium with a Gaussian profile,
\begin{equation}
    \rho_\mathrm{dust, i}(r,z) = \frac{\Sigma_\mathrm{dust, i}(r)}{\sqrt{2\pi} H_\mathrm{dust,i}} \exp\left(-\frac{z^2}{2H_\mathrm{dust,i}^2}\right),
\end{equation}
where $\Sigma_\mathrm{dust,i}(r)$ is the surface density of the $i$th grain size species in the dust evolution simulations and $H_\mathrm{dust,i}$ is the dust scale height from \citet{dubrulle1995}. We show the dust vertical distributions of different grain sizes for the varying-$\vf$ model at 1~Myr in Fig.~\ref{fig:radmc_setup_1myr} as an example. Small grains ($<5\,\mu$m) are well coupled to the gas. Medium grains (5--100~$\mu$m) show strong depletion in the gap region and are partially settled to the midplane. Grains larger than 100~$\mu$m are settled to the midplane, and their outer edge indicates inward drift.

When making synthetic observations, we use the 2D $(r,\theta)$ temperature structure (last panel of Fig.~\ref{fig:radmc_setup_1myr}) from the $3\,\Mj$ planet at 30~au model in \citet{chen2024}, obtained by iterating hydrodynamical and radiative transfer simulations to reach a self-consistent temperature and gas density structure. The midplane temperature in this 2D cut matches the 1D midplane profile in Fig.~\ref{fig:setup_visc1e-3}. We do not use $\radmc$ to recalculate the temperature structure for the following reason: our dust evolution domain starts at 5~au, whereas the full \citet{chen2024} disk extends from 1 to 100~au. If we reran thermal RT with an inner boundary at 5~au, $\radmc$ would treat the disk as having an inner cavity and produce temperatures much higher than in our setup. For self-consistency, we import the temperature structure from the iterative simulations for all synthetic observations.

We set the distance to 140~pc and assume a face-on disk. Images are produced with \texttt{radmc3d image} at 1.3~mm, 13~$\mu$m, and 2~$\mu$m. These wavelengths correspond to ALMA (1.3~mm), upcoming ELT/METIS (13~$\mu$m), and SPHERE/VLT (2~$\mu$m). At 1.3~mm, the emission is dominated by thermal ray tracing from the imported dust temperature field. At 2~$\mu$m, the emission is dominated by scattered light. At 13~$\mu$m, the emission is a mixture of thermal and scattered light: thermal-dominated near the ice-line rings (25--36~au), but scattered-light-dominated beyond $\sim$50~au. The 13 and 2~$\mu$m emission thus comes from above the midplane, whereas we evolve the dust only in 1D and extend it vertically with the Gaussian profile above. Because vertical mixing, on a timescale $1/(\alpha\Omega)\sim2\times10^4$~yr at 25~au, is much faster than the Myr dust evolution, we expect the 13 and 2~$\mu$m emission from the disk surface still gives the same ring positions as our 1D dust evolution model. We use $\nphotsca=10^6$ and isotropic scattering in making the ray-tracing image. We convolve the images and radial intensity profiles with a Gaussian beam of 0.05~arcsec, comparable to the resolution of ALMA and SPHERE.

\subsection{Radial intensity profiles}
\label{sec:rt-intensity}

\begin{figure*}
	\includegraphics[width=\linewidth]{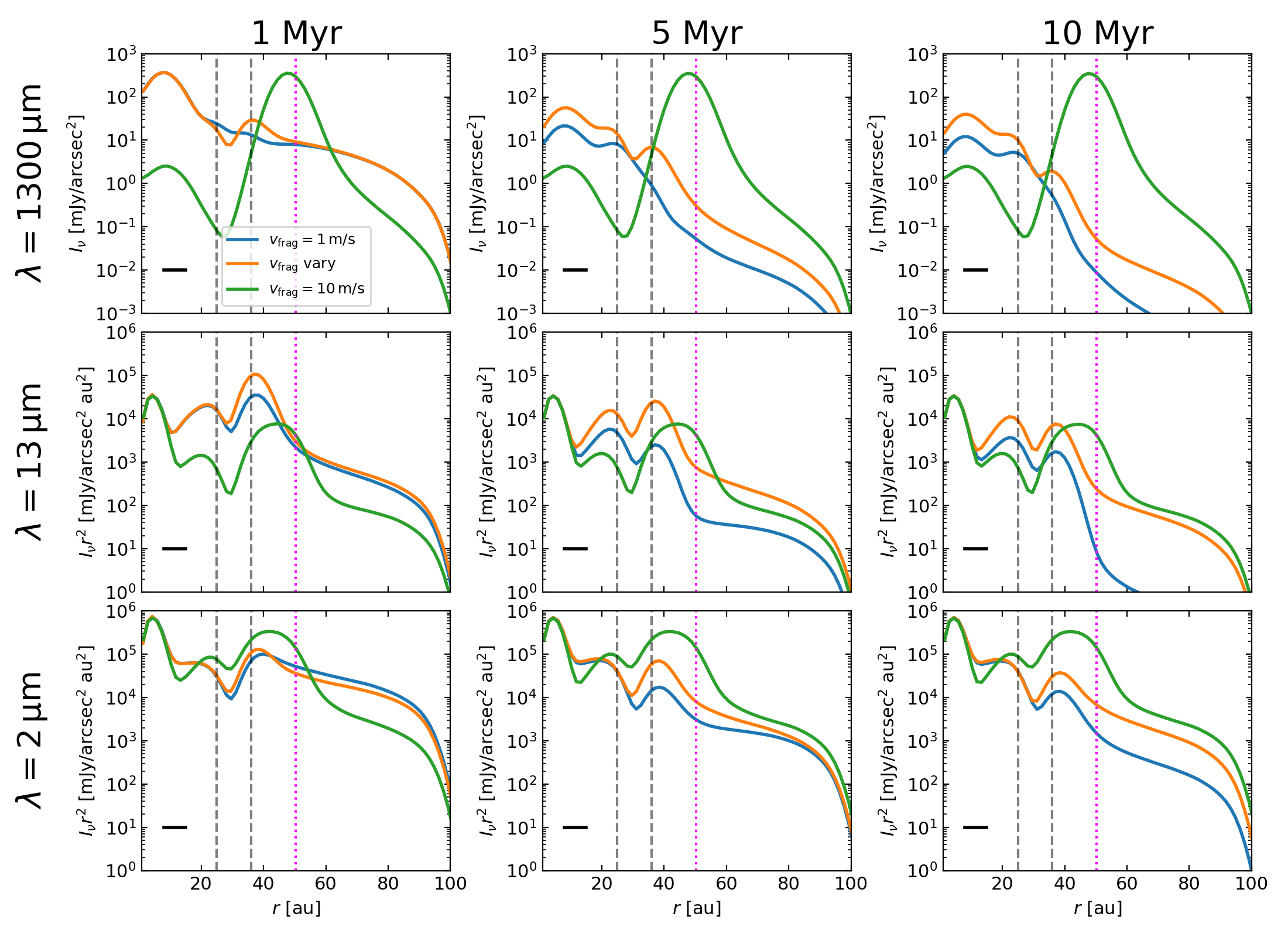}
    \caption{
    Radial intensity profiles at 1~Myr (left), 5~Myr (middle), and 10~Myr (right) for the three $\vf$ models (blue: $\vf=1$~m/s; orange: varying-$\vf$; green: $\vf=10$~m/s) in a $3\,\Mj$ planet at 30~au disk with $\alpha=10^{-3}$. Rows show 1.3~mm continuum (top), 13~$\mu$m emission (thermal and scattered light; middle), and 2~$\mu$m scattered light (bottom). Profiles are convolved with a 0.05~arcsec Gaussian beam (scale bar in each panel). Vertical grey dashed lines mark the \coo\ ice lines near 25 and 35~au, and the magenta dotted line marks the location of the gas pressure maximum (pressure bump) at the outer gap edge, as in Fig.~\ref{fig:sdust_visc1e-3}. For 2 and 13~$\mu$m, $I_\nu r^2$ is plotted to compensate for the $1/r^2$ dilution of stellar illumination for the scattered light image and to highlight disk substructure.
    }
    \label{fig:int_beam}
\end{figure*}

We show radial intensity profiles of 1.3~mm, 13~$\mu$m, and 2~$\mu$m for all three models at 1, 5, and 10~Myr in Fig.~\ref{fig:int_beam} (columns: time; rows: wavelength; coloured curves: $\vf$ model; see caption). The corresponding beam-convolved synthetic images are in Appendix Figs.~\ref{fig:images_1300um}, \ref{fig:images_13um}, and \ref{fig:images_2um}. For 2 and 13~$\mu$m, the plotted quantity is $I_\nu r^2$ (see Fig.~\ref{fig:int_beam} caption).

Overall, the varying-$\vf$ model (orange curves) shows the clearest and most persistent ice-line-related substructure. In the gap, the profiles often exhibit two features tied to the two \coo\ ice lines (near 25 and 35~au): either two distinct peaks or, at minimum, a shoulder at the inner ice line and a ring at the outer ice line. This morphology is most pronounced in the middle column (5~Myr) and is not mimicked by the $\vf=10$~m/s model (green curves), which is instead dominated by a single bright ring just interior to the pressure bump (magenta dotted line, $\sim$50~au), consistent with the $\amax$ and $\sigmad$ peaks sitting slightly inside the pressure maximum in Fig.~\ref{fig:sdust_visc1e-3}. The $\vf=1$~m/s model (blue curves) can also show paired features at 13 and 2~$\mu$m, but its ice-line contrast is weaker, its outer disk fades more strongly with time, and its 1.3~mm continuum is smoother in the gap.

We now quantify how these features evolve with time (columns of Fig.~\ref{fig:int_beam}).
At 1~Myr (left column), the 1.3~mm continuum is similar in the varying-$\vf$ and $\vf=1$~m/s models, but near the \coo\ ice lines the varying-$\vf$ model already shows a ring-to-gap contrast of $\sim$3, whereas the $\vf=1$~m/s gap is nearly smooth. The $\vf=10$~m/s model instead forms a strong trap ring at the outer gap edge ($\sim$50~au), producing a transition-disk-like inner cavity at 1.3~mm (Fig.~\ref{fig:images_1300um}) with a ring-to-gap contrast $\gtrsim10^{3}$ at all three epochs. At 13 and 2~$\mu$m, both the varying-$\vf$ and $\vf=1$~m/s models show two ice-line peaks (widths $\sim$10~au at 13~$\mu$m), while the $\vf=10$~m/s trap ring dominates and shifts inward with decreasing wavelength, from $\sim$50~au at 1.3~mm to $\sim$45~au at 13~$\mu$m and $\sim$40~au at 2~$\mu$m, as larger grains are trapped farther out. At 13~$\mu$m the outer ice-line ring is $\sim$2--3$\times$ brighter than the inner one in the varying-$\vf$ model (the inner line appearing mainly as a shoulder), whereas the two are comparable for $\vf=1$~m/s.

At 5~Myr (middle column), the $\vf=10$~m/s profiles are nearly unchanged. The $\vf=1$~m/s model has lost much of its outer-disk 1.3~mm emission (fainter by two order of magnitude than at 1~Myr), leaving only faint paired features at 13~$\mu$m. The varying-$\vf$ model shows its clearest ice-line signature: two well-defined 13~$\mu$m rings of comparable brightness (ratio $\sim$1, widths $\sim$10~au), and a prominent outer ice-line ring with an inner shoulder at 1.3~mm and 2~$\mu$m.

At 10~Myr (right column), the $\vf=10$~m/s model remains stable. Both the $\vf=1$~m/s and varying-$\vf$ models keep depleting the outer disk while preserving their 5~Myr inner-gap structure; the outer 13~$\mu$m ring in the varying-$\vf$ model now fades to slightly below the inner ring. The varying-$\vf$ model stays brighter than $\vf=1$~m/s in the inner disk and at 13~$\mu$m (by a factor of $\sim$2--3), reflecting its greater retention of large grains interior to the ice lines.

\section{Discussion}
\label{sec:discussion}

\subsection{Comments on recent or future observations}
\label{sec:discussion-obs}

ALMA surveys find no simple match between dust ring locations and ice-line radii inferred from smooth, luminosity-based temperature profiles \citep{huang2018,long2018}. Our models suggest one reason: a planet-opened gap can create a non-smooth temperature structure, shifting the \coo\ ice lines to $\sim$25 and $\sim$35~au (Fig.~\ref{fig:setup_visc1e-3}). Ice-line-related substructure should therefore not necessarily coincide with ice-line locations predicted from smooth disk models.

To place our models in context, we compare our 1.3~mm ring-to-gap contrast (first row of Fig.~\ref{fig:int_beam}) with DSHARP gaps at 20--40~au, around the radius of our model gap \citep{huang2018}. Seven of the eight such gaps with measured depths are shallow, with ring-to-gap ratios of $\sim$1.0--1.3, while only HD~143006 D22 reaches $\sim$25 (Table~1 in \citet{huang2018}). Our varying-$\vf$ model falls in the shallow class. Its modest millimetre ring-to-gap contrast at the \coo\ ice lines (Section~\ref{sec:rt-intensity}) suggests that low-contrast millimetre rings can arise from ice-line-induced $\vf$ variations alone, without strong dust trapping. Around brighter stars, the \coo\ ice lines, and hence such rings, would lie farther out. The deep, high-contrast rings instead require efficient trapping at a gas pressure maximum, as in our $\vf=10$~m/s model. That model, however, becomes optically thick ($\tau_{\rm abs}\sim1$) and overshoots the observed contrasts, which remain marginally optically thin ($\tau_{\rm abs}\approx0.2$--$0.5$; \citealt{dullemond2018}); we therefore treat it as an illustration of the trapping mechanism rather than a match to any individual source.

The clearest observational prediction of the varying-$\vf$ model is paired
structure near the two \coo\ ice lines in 13~$\mu$m emission, especially around
5~Myr.
This motivates a multi-wavelength test combining ELT mid-infrared imaging with
ALMA 1.3~mm continuum and shorter-wavelength scattered-light observations, which
separates the two scenarios by how the ring position depends on wavelength.
At 140~pc, the ELT diffraction limit at 13~$\mu$m is $\sim$0.07~arcsec ($\sim$10~au), at which the paired features at 5~Myr remain separated.
Because ice-line rings are set by temperature, they appear at the same radii
($\sim$25 and $\sim$35~au) from 2~$\mu$m to 1.3~mm; paired features whose peaks
stay fixed across wavelength would support the hot-gap ice-line scenario.
A dust trap, by contrast, sorts grains by size and forms a single ring that
dominates the disk, whose intensity peak moves closer to the gap at shorter
wavelengths (from $\sim$50~au at 1.3~mm to $\sim$40~au at 2~$\mu$m in our
$\vf=10$~m/s model); such a wavelength-dependent ring would instead favour
pressure-bump trapping.

Beyond this multi-wavelength dust test, the hot-gap origin itself can be checked by combining dust and gas observations. \citet{chen2025} suggests that a planet-opened gas gap can appear as a molecular ring because volatile ices are sublimated in the locally heated gap. In this work, we find that the same hot-gap structure can also produce paired dust rings at MIR through ice-line-induced changes in $\vf$. Spatially coincident molecular emission rings and dust MIR substructure would therefore provide a multi-tracer test of the hot-gap scenario. At 140~pc, a molecular ring filling the $\sim$27~au wide gap spans $\sim$0.2~arcsec, about twice the $\sim$0.1~arcsec resolution targeted by recent ALMA line surveys \citep{teague2025}, so resolving it requires deep, high-resolution line observations. This comparison should be made with caution, however, because the hot-gap ice lines in our model are set mainly by the midplane temperature, whereas molecular emission may trace warmer disk surface layers. The dust rings are less affected, because vertical mixing carries their midplane structure to the surface layers (Section~\ref{sec:rt-setup}).

\subsection{Conditions for observable ice-line rings}
\label{sec:discussion-conditions}

We mainly discuss the case of a $3\,\Mj$ planet at 30~au with $\alpha=10^{-3}$ in Sections~\ref{sec:dust-results} and \ref{sec:rt-intensity}, but one may ask how commonly such ice-line rings can be observed in disks with different planet and disk properties.
Here we set out which planet and disk properties allow the ice-line rings
to form, and under what observational conditions they can be detected.

Both the planet mass and its orbital radius matter for opening a gap hot enough
to sublimate \coo. The criterion is that the gap midplane temperature exceeds
47~K. The planet location sets the background temperature, which is 27~K at
30~au and 32~K at 10~au without a planet. The gap depth sets how far the gap is
heated above it (Table~\ref{tab:gapT}). At 30~au, the $3\,\Mj$ gap is depleted
to $\sim$$10^{-3}$ and reaches 51~K, whereas the shallower $1\,\Mj$ gap
($\sim$$4\times10^{-2}$) reaches only 39~K. Restricting attention to the region
outside $\sim$10~au that current observations resolve, and to a typical disk
around a solar-mass star, Figs.~D1 and D2 in \citet{chen2024} show that the gap
becomes hot enough to sublimate \coo\ ice for a $1\,\Mj$ planet at 10~au, a
$3\,\Mj$ planet at 10~au, and a $3\,\Mj$ planet at 30~au. The smaller aspect
ratio at 10~au makes the gap easier to open. We have no models with planets
beyond 30~au, where the disk is colder and a gap needs more heating to sublimate
\coo. The temperature deviations beyond 50~au found by \citet{izquierdo2023},
however, are in disks around Herbig stars (MWC~480 and HD~163296), whose warmer
disks could place the \coo\ ice lines, and hence such rings, at larger radii.

The second condition is that the two ice lines lie far enough apart to be
resolved. For the $3\,\Mj$, 30~au model they are $\sim$10~au apart, which a
0.05~arcsec beam at 140~pc (7~au) resolves. For planets at 10~au, the ice lines
are only $\sim$3--7~au apart (Table~\ref{tab:gapT}), below the $\sim$10~au ELT
resolution at 13~$\mu$m (Section~\ref{sec:discussion-obs}), so any paired rings
there would be unresolved.

Within this disk, the ring positions do not depend on the size of the $\vf$
jump. Lowering $\vf$ between the ice lines from 10 to 6~m/s leaves the rings at
the same radii and only lowers their contrast (Appendix~\ref{sec:app-vfcontrast}).

\subsection{Pebble flux from different $\vf$ models}
\label{sec:discussion-pebble}

\begin{figure}
    \includegraphics[width=\columnwidth]{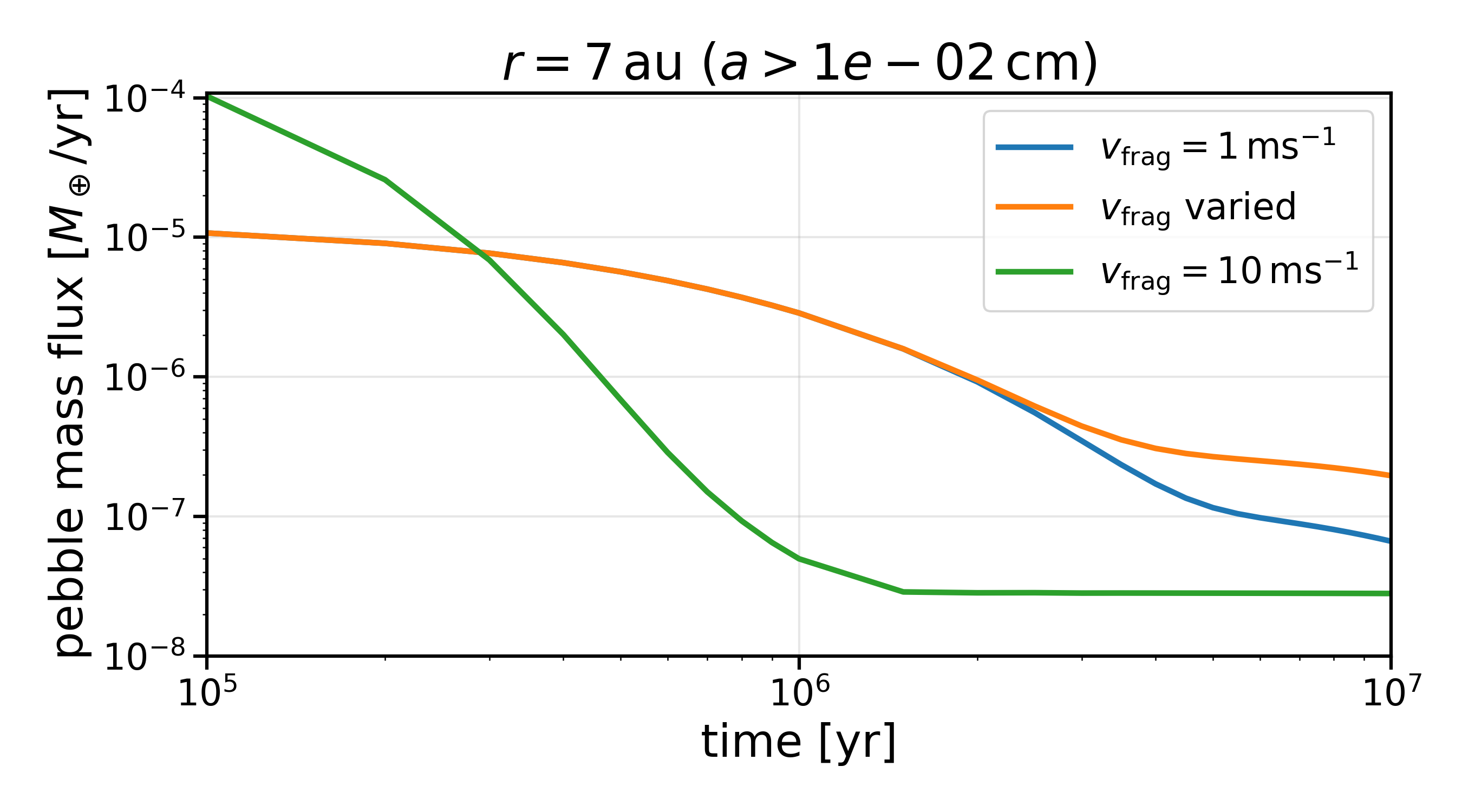}
    \caption{Inward pebble mass flux at $r=7$~au as a function of time for $\vf=1$~m/s (blue), varying-$\vf$ (orange), and $\vf=10$~m/s (green). Only grains with $a>0.01$~cm are included.
}
\label{fig:pebble_flux}
\end{figure}

The impact of the pebble flux in the inner disk emission is one of the observable quantities, for example, resupply of pebbles from beyond the water ice line can help sustain the inner-disk water vapour excess detected by JWST \citep{banzatti2023}.
We show the inward pebble flux at $r=7$~au for the three models in Fig.~\ref{fig:pebble_flux}. At each radius, we define the pebble mass flux as
\begin{equation}
    \dot{M}(r) = 2\pi r \sum_{i} \Sigma_{\mathrm{dust},i}(r)\, v_{r,i}(r),
    \label{eq:pebble_flux}
\end{equation}
where the sum runs over all grain-size species $i$ with $a_i \geq a_{\mathrm{peb}} = 0.01$~cm, $\Sigma_{\mathrm{dust},i}$ is the dust surface density of species $i$, and $v_{r,i}$ is its radial velocity. We evaluate Eq.~\eqref{eq:pebble_flux} at $r=7$~au and plot $-\dot{M}$ so that inward transport is positive.

For $\vf=10$~m/s, the inward flux is initially the highest, but it declines most rapidly and reaches a low plateau by $\sim1.5\times10^6$~yr. This behaviour is consistent with efficient grain growth at the beginning followed by strong dust trapping at the pressure bump outside the planet-opened gap (at $\sim50$~au), which accumulates pebble-sized grains in the outer disk and largely shuts off their radial drift into the inner disk. In contrast, the $\vf=1$~m/s model shows a lower but more gradual decline, as frequent fragmentation keeps grains smaller and maintains a steadier, weaker inward delivery. The varying-$\vf$ model closely follows the $\vf=1$~m/s model before $\sim$2.5~Myr, but sustains a higher inward flux after that, indicating that radial structure in $\vf(r)$ can prolong pebble supply to the inner disk relative to a uniform low fragmentation velocity.

\citet{banzatti2023} found excess cool water in compact disks compared with large disks with gaps, and attributed it to gaps that block pebble drift. In our models, only the $\vf=10$~m/s gap blocks the pebbles, reducing the flux at $\sim$1~Myr by two orders of magnitude. The varying-$\vf$ model delivers pebbles as efficiently as the $\vf=1$~m/s model, even though its gap produces ice-line rings. A disk with a leaky gap may therefore still show an inner-disk water excess, unlike a disk with a dust-trapping gap. In our models the same deep gap is leaky or trapping depending on $\vf$ \citep[see also][]{kalyaan2023}. \citet{banzatti2023} also note that compact disks may host unresolved gaps, and that shallow gaps can be leaky enough to barely reduce the inner-disk water enrichment.

\subsection{Limits of this work}

We do not include grain-size changes from volatile sublimation or condensation when grains drift across ice lines (the traffic-jam mechanism of \citet{drazkowska2017}). Our model instead includes only ice-line-induced jumps in $\vf$ \citep{pinilla2017}. In our domain (5--100~au), the relevant transitions are \coo\ ice lines. Because \hho\ makes up roughly half the mass (and $\sim$75\% of the volume) of icy grains beyond its snowline \citep{drazkowska2017}, whereas \coo\ ice is only $\sim$30\% as abundant as water \citep{oberg2011}, volatile exchange at \coo\ lines would modify grain sizes far less than at the water ice line. The grain-size jumps we see at \coo\ lines therefore reflect $\vf$ changes, not mass and grain-size exchange across the ice line.

We implement ice-line phase transitions as abrupt jumps in $\vf(r)$, whereas in reality sublimation and condensation occur over finite radial (and vertical) zones. We therefore likely overestimate the sharpness and contrast of the grain-size and $\sigmad$ peaks near 25 and 35~au, and may exaggerate how distinct the two ice-line rings are. A full treatment of phase-change physics in the radial and vertical directions to model the ice line locations \citep{wang2025} is beyond our scope.

Our 1D model of the gas and dust substructure simplifies the 2D and 3D effects of the planet, such as vertical dust diffusion and spirals. In 3D, a Jupiter-mass planet stirs mm-sized dust vertically near its gap, raising the effective vertical diffusivity to $\sim$$5\times10^{-3}$--$2\times10^{-2}$ \citep{binkert2023}, well above our $\alpha=10^{-3}$. This would lower the midplane dust-to-gas ratio, and hence planetesimal formation at the ice-line peaks, but would not move the peaks, whose positions are set by the temperature. Shock heating by the spirals \citep{ziampras2020} would further raise the gap temperature. This would allow lower-mass planets to heat their gaps above the \coo\ sublimation temperature, and would move the two \coo\ ice lines farther apart, widening the separation between the rings.

We assume isotropic scattering for the synthetic scattered-light images and intensity profiles. More realistic scattered-light images would require anisotropic scattering. Our study focuses on azimuthally averaged radial profiles, which are dominated by surface density, grain size distribution, and disk geometry (scale height, flaring). Detailed angular redistribution of light mainly affects azimuthal brightness and near/far-side contrast, so isotropic scattering is sufficient for the qualitative shape of the azimuthally averaged profiles.

\section{Conclusions}

We study how a hot, planet-opened gap affects dust evolution and its observational signatures. We adopt fixed gas and temperature profiles from the \citet{chen2024} model of a $3\,\Mj$ planet at 30~au with $\alpha=10^{-3}$, in which the heated gap sublimates \coo\ ice between two ice lines near 25 and 35~au (Fig.~\ref{fig:setup_visc1e-3}). We compare a varying-$\vf$ model, in which $\vf$ rises from 1 to 10~m/s where grains lose their \coo\ ice and are coated only with water ice (Fig.~\ref{fig:vfrag_r}), with uniform $\vf=1$ and 10~m/s models over 1--10~Myr.

When $\vf$ varies across ice lines, dust accumulates with enhanced grain sizes at the two \coo\ ice lines near 25 and 35~au (Section~\ref{sec:dust-results}). This ice-line substructure persists over 1--10~Myr and retains more outer-disk dust than a uniform $\vf=1$~m/s model, which instead depletes the outer disk through drift, whereas a uniform $\vf=10$~m/s model produces a strong dust trap at the pressure bump near 50~au rather than paired ice-line features. Dust trapping also governs the pebble supply to the inner disk: the strongly trapped $\vf=10$~m/s model suppresses the inward pebble flux by up to two orders of magnitude relative to the varying- and $\vf=1$~m/s models, so an inner-disk volatile excess would trace the trapping strength (Section~\ref{sec:discussion-pebble}).

In synthetic observations (Section~\ref{sec:rt-intensity}), the varying-$\vf$ model shows the clearest ice-line signature as paired rings in 13~$\mu$m emission around 5~Myr, while at 1.3~mm it produces only a modest ring-to-gap contrast (a factor of a few), consistent with the shallow, low-contrast rings that make up a large fraction of the DSHARP survey (Section~\ref{sec:discussion-obs}). The $\vf=10$~m/s model instead yields a single, high-contrast ring at the pressure bump, representing the deep, high-contrast rings seen in a minority of DSHARP disks that require efficient dust trapping.

These signatures motivate a multi-wavelength test (Section~\ref{sec:discussion-obs}): ice-line rings stay at fixed radii ($\sim$25 and $\sim$35~au) from scattered light to the millimetre, whereas a pressure-bump trap forms a single dominant ring whose peak migrates inward at shorter wavelengths. Resolving paired, wavelength-independent features would thus support ice-line-induced dust structure in a hot, planet-opened gap, while a single, wavelength-dependent ring would favour dust trapping.

\section*{Acknowledgements}

We are thankful to the referee for the constructive report. We thank Lilian Luo, Ted Bergin and Andrew Allan for helpful discussions. This work used the Hypatia and Mentat computing clusters at University College London. P.P. acknowledges funding from the UK Research and Innovation (UKRI) under the UK government's Horizon Europe funding guarantee from ERC (under grant agreement No 101076489).

\bibliographystyle{aasjournalv7}
\bibliography{overleaf_dustpy.bib}

\appendix

\section{Synthetic images}

\begin{figure*}
	\includegraphics[width=\linewidth]{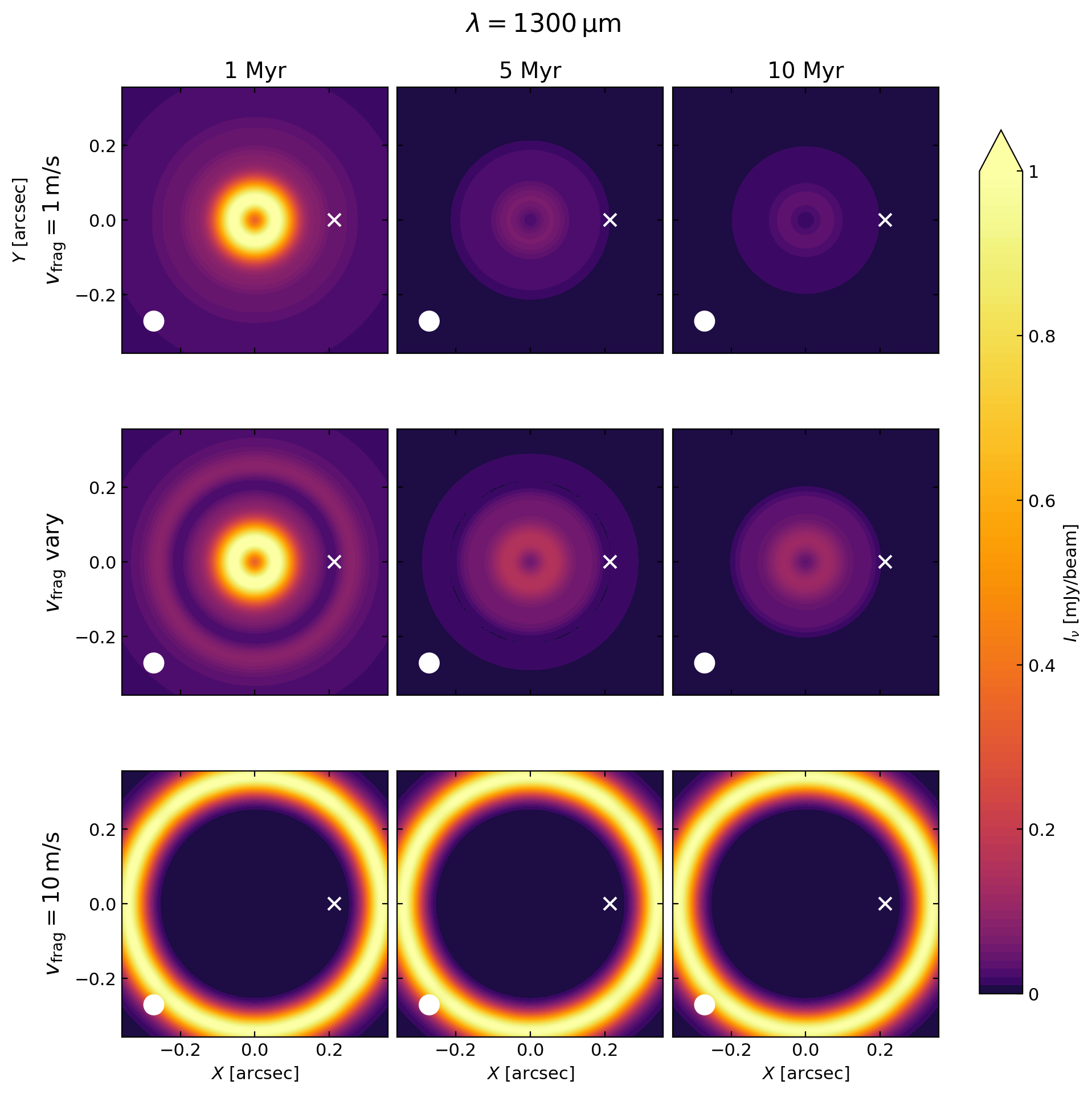}
    \caption{Synthetic 1.3~mm continuum images convolved with a
    0.05~arcsec Gaussian beam. Rows show $\vf=1$~m/s (top),
    varying-$\vf$ (middle), and fixed
    $\vf=10$~m/s (bottom); columns show 1, 5, and 10~Myr of dust
    evolution (left to right). The intensity scale is $I_\nu$ in
    mJy~beam$^{-1}$.}
    \label{fig:images_1300um}
\end{figure*}

\begin{figure*}
	\includegraphics[width=\linewidth]{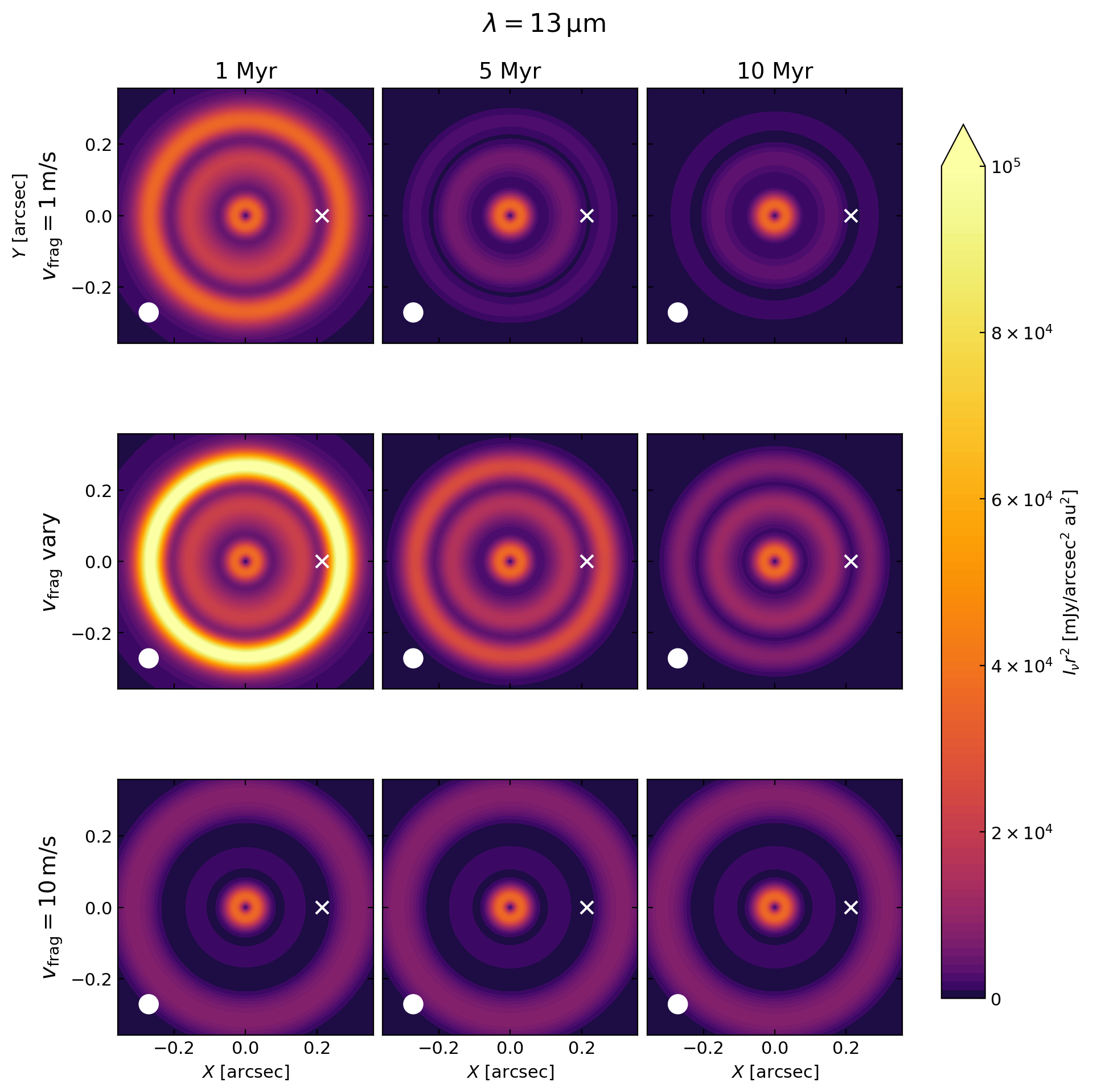}
    \caption{Same as Fig.~\ref{fig:images_1300um}, but for emission (thermal and scattered light) at
    13~$\mu$m. The displayed quantity is $I_\nu r^2$ in
    mJy~arcsec$^{-2}$~au$^2$.}
    \label{fig:images_13um}
\end{figure*}

\begin{figure*}
	\includegraphics[width=\linewidth]{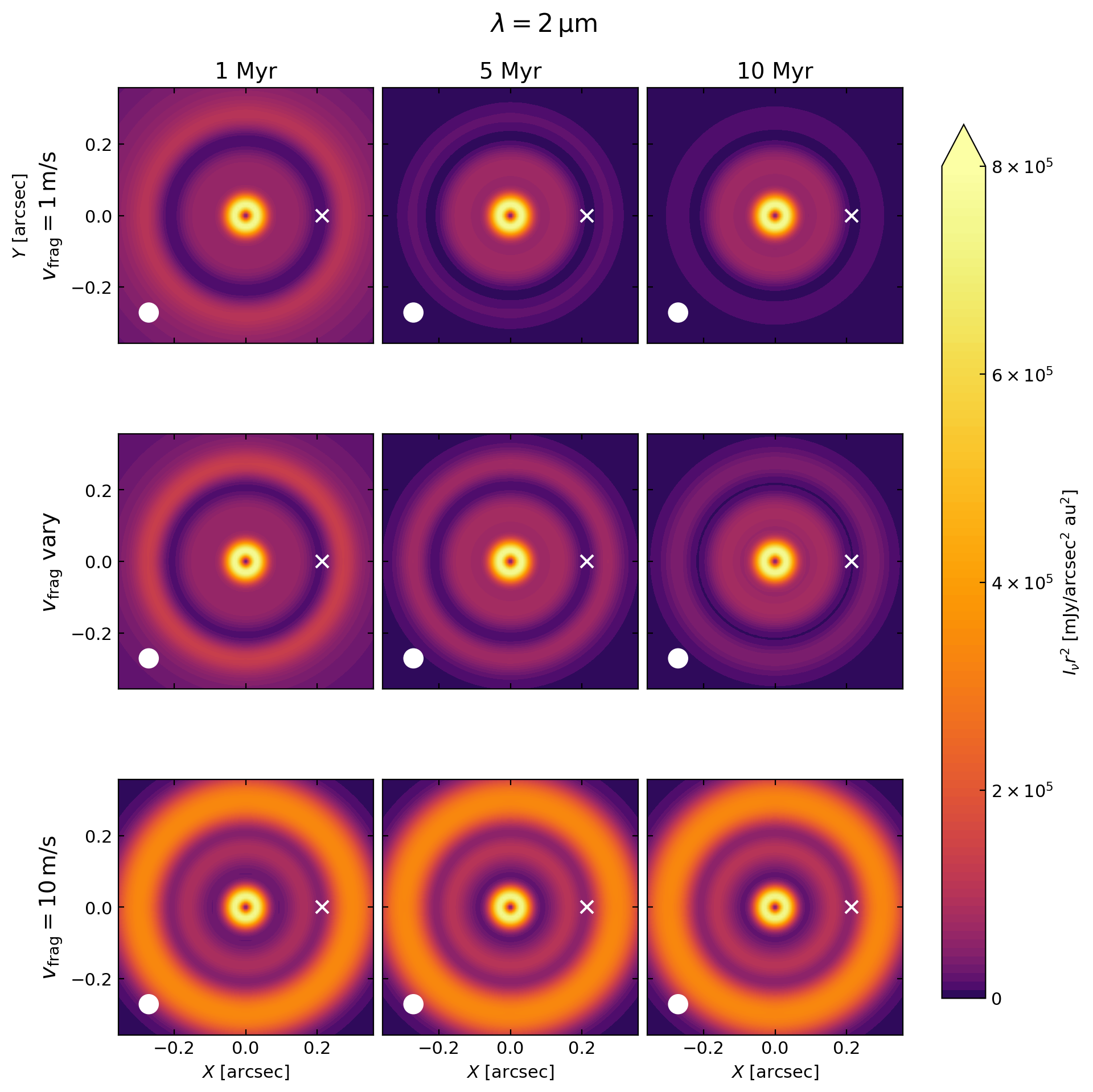}
    \caption{Same as Fig.~\ref{fig:images_1300um}, but for scattered light at
    2~$\mu$m. The displayed quantity is $I_\nu r^2$ in
    mJy~arcsec$^{-2}$~au$^2$.}
    \label{fig:images_2um}
\end{figure*}

We show beam-convolved synthetic images at 1.3~mm, 13~$\mu$m, and 2~$\mu$m in
Figs.~\ref{fig:images_1300um}, \ref{fig:images_13um}, and \ref{fig:images_2um},
respectively. Each figure compares the three $\vf$ assumptions (rows) at
1, 5, and 10~Myr (columns) for a $3\,\Mj$ planet at 30~au in an
$\alpha=10^{-3}$ disk at 140~pc. The planet location is marked by a white
cross; the white circle in each panel indicates the 0.05~arcsec Gaussian beam.

At 1.3~mm (Fig.~\ref{fig:images_1300um}), the varying-$\vf$ shows modest ring-gap contrast at the two \coo\ ice line region
and $\vf=1$~m/s models show almost no substructure in the gap, consistent with the radial profiles in Fig.~\ref{fig:int_beam}. The $\vf=10$~m/s model instead produces a bright ring at the pressure bump near $\sim$50~au that changes little with time.

At 13~$\mu$m (Fig.~\ref{fig:images_13um}), the varying-$\vf$ model shows the clearest ice-line signature as paired rings near the two \coo\ ice lines in the gap, especially around 5~Myr.

At 2~$\mu$m (Fig.~\ref{fig:images_2um}), the varying-$\vf$ and $\vf=1$~m/s models show a shoulder near the inner \coo\ ice line together with an outer ring near the outer ice line in the gap region.

\section{Gap temperature and \coo\ ice lines across the hydrodynamical models}
\label{sec:app-gapT}

\begin{figure*}
    \includegraphics[width=\linewidth]{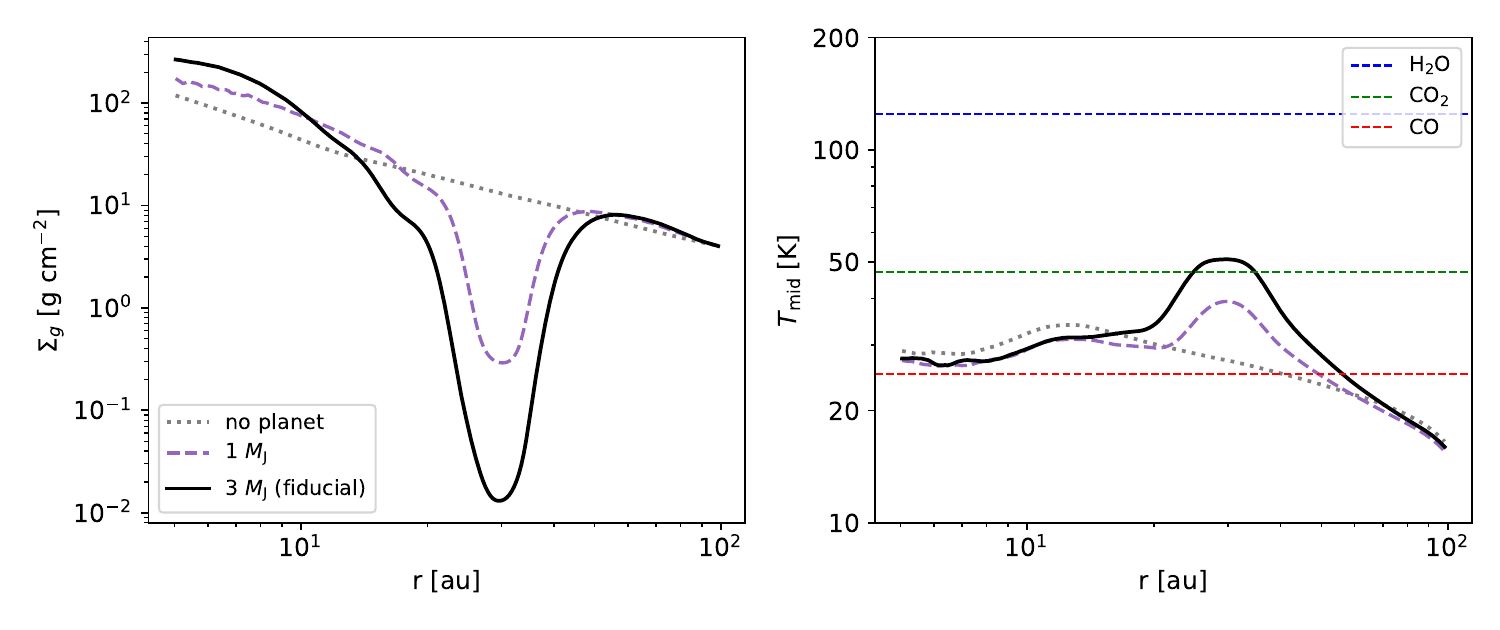}
    \caption{Gas surface density (left) and midplane temperature (right) of the no-planet, $1\,\Mj$ and fiducial $3\,\Mj$ models of \citet{chen2024}, all with the planet at 30~au. Horizontal dashed lines mark the \hho, \coo\ and \co\ sublimation temperatures.}
    \label{fig:refmodels}
\end{figure*}

Table~\ref{tab:gapT} and Fig.~\ref{fig:refmodels} compare the gap temperature and \coo\ ice lines in the \citet{chen2024} models. We use their $10\,M_\oplus$ model to represent the no-planet case, since it opens no gap and does not change the disk temperature. Without a planet, the midplane temperature decreases smoothly and stays below the \coo\ sublimation temperature throughout our grid. The $1\,\Mj$ gap reaches only 39~K, so \coo\ stays frozen and $\vf=1$~m/s everywhere. A varying-$\vf$ model would therefore be identical to the uniform $\vf=1$~m/s model, with no ice-line rings. In all three models \co\ is already sublimated inside $\sim$40~au, and its ice line does not change $\vf$. We adopt the $3\,\Mj$, 30~au model because it is the only one whose gap exceeds the \coo\ sublimation temperature and whose two ice lines lie far enough apart to be resolved (Section~\ref{sec:discussion-conditions}).

\begin{deluxetable*}{lccccc}
\tablecaption{Gap temperature and \coo\ ice lines in the \citet{chen2024}
models. $\gd$ is measured at $\rp$ relative to $2\,\rp$. $T_\mathrm{bg}$ is the
midplane temperature at $\rp$ in the no-planet model. The last two columns give
the \coo\ ice-line radii and their separation.\label{tab:gapT}}
\tablehead{
\colhead{Model} & \colhead{$T(\rp)$} & \colhead{$T_\mathrm{bg}$} &
\colhead{$\gd$} & \colhead{\coo\ lines} & \colhead{$\Delta r$} \\
\colhead{} & \colhead{(K)} & \colhead{(K)} & \colhead{} & \colhead{(au)} &
\colhead{(au)}
}
\startdata
no planet ($10\,M_\oplus$, 30~au) & 27.2 & 27.2 & 2.0 & none & \nodata \\
$1\,\Mj$, 30~au & 39.2 & 27.2 & $3.8\times10^{-2}$ & none & \nodata \\
$3\,\Mj$, 30~au (fiducial) & 50.9 & 27.2 & $1.7\times10^{-3}$ & 25, 35 & 10.1 \\
$3\,\Mj$, 10~au & 61.2 & 32.2 & $9.8\times10^{-4}$ & 7.4, 14.5 & 7.1 \\
$1\,\Mj$, 10~au & 61.1 & 32.2 & $2.6\times10^{-4}$ & 8.3, 11.6 & 3.4 \\
\enddata
\end{deluxetable*}
\onecolumngrid   %

\section{Sensitivity to the fragmentation-velocity contrast}
\label{sec:app-vfcontrast}

\begin{figure*}
    \includegraphics[width=\linewidth]{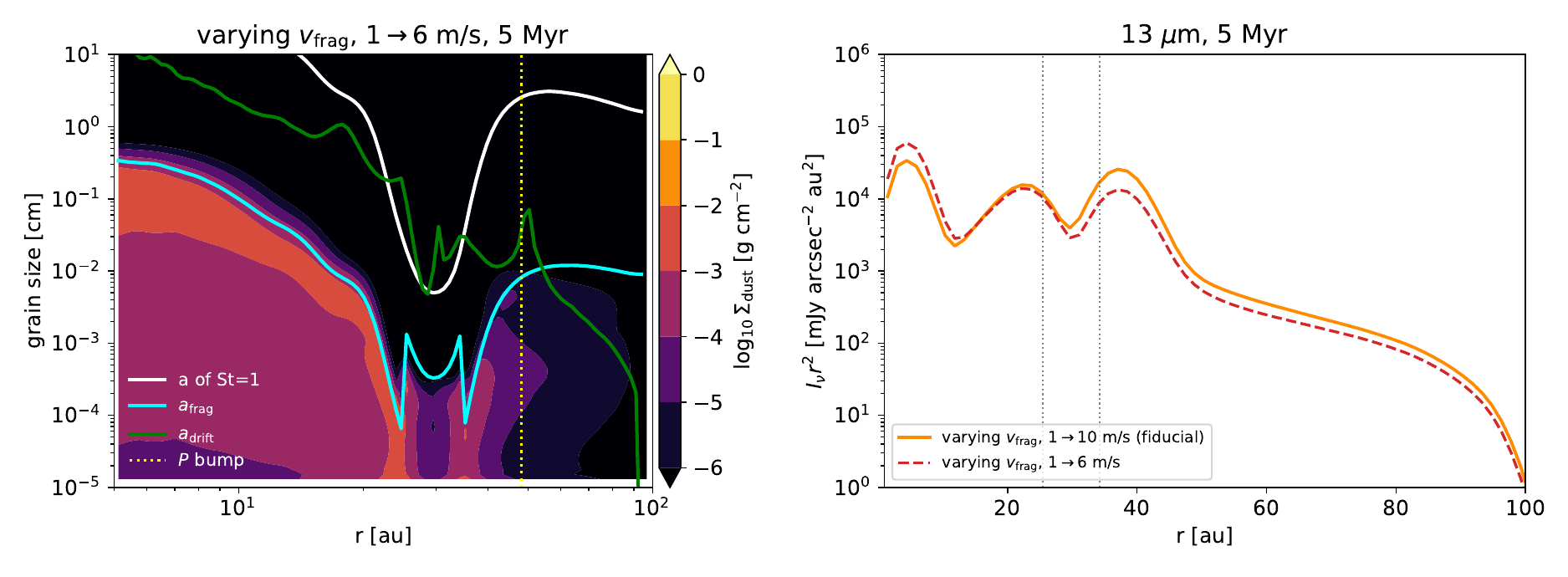}
    \caption{Varying-$\vf$ model with $\vf$ rising from 1 to 6~m/s, instead of 1 to 10~m/s, between the \coo\ ice lines, at 5~Myr. Left: dust size distribution, as in Fig.~\ref{fig:sdust_visc1e-3}. Right: 13~$\mu$m intensity profiles for the varying-$\vf$ models with 1$\rightarrow$10~m/s (orange) and 1$\rightarrow$6~m/s (red dashed), convolved with the 0.05~arcsec beam as in Fig.~\ref{fig:int_beam}. Dotted lines mark the \coo\ ice lines.}
    \label{fig:vfcontrast}
\end{figure*}

Laboratory results for grains coated only with water ice range from a sticking threshold of $\sim$10~m/s \citep{gundlach2014} to a much lower stickiness below $\sim$175~K \citep{musiolik2019}. We therefore repeat the varying-$\vf$ model with an intermediate value of 6~m/s between the \coo\ ice lines (Fig.~\ref{fig:vfcontrast}). In the 10~m/s model, $\amax$ between the ice lines reaches only $\sim$20--50\% of $\afrag$, so lowering $\vf$ to 6~m/s reduces $\afrag$ by a factor of $\sim$3 but $\amax$ by only $\sim$30\%. The dust still piles up at the two \coo\ ice lines, and the paired 13~$\mu$m rings stay at the same radii. The outer ring is about half as bright as in the 10~m/s model.

\end{document}